\documentclass[aps,prl
]{revtex4}
\def\frac#1#2{{\textstyle{#1\over#2}}} 

\def\ket#1{| #1\rangle}

\def\R{\hbox{\rm I \kern-5pt R}}

\usepackage{amssymb}
\usepackage{amsmath}
\usepackage[sort&compress]{natbib}
\usepackage[dvips]{graphicx}

\begin{document}


\title{Why Classical Certification is Impossible in a Quantum World}
\author{ Adrian Kent}
\email{a.p.a.kent@damtp.cam.ac.uk}
\affiliation{Centre for Quantum Information and Foundations,
Department of Applied Mathematics and
Theoretical Physics, University of Cambridge, U.K.}
\affiliation{Perimeter Institute for Theoretical Physics, Waterloo, Ontario, Canada}
\date{September 2004 (revised April 2011, discussion and crypto model
  extended, refs updated)}
\begin{abstract}
We give a simple proof that it is impossible to guarantee
the classicality of inputs into any mistrustful 
quantum cryptographic protocol.  The argument 
illuminates the impossibility of unconditionally secure 
quantum implementations of essentially classical tasks such as bit
commitment with a certified classical committed bit, 
classical oblivious transfer, and secure classical
multi-party computations of secret classical data.  
It applies to both non-relativistic and relativistic protocols.   
\end{abstract}
\maketitle
\section{Introduction} 
Wiesner's pioneering work in quantum cryptography \cite{wiesner}, 
and the ensuing discoveries by Bennett and
Brassard of secure quantum key distribution \cite{BBeightyfour}
and by Ekert of entanglement-based quantum key distribution
\cite{ekert}, created much interest in the possibility of secure quantum
implementations of other cryptographic tasks. 
In particular, there has recently been a great deal of interest in exploring
quantum implementations of cryptographic tasks involving
mutually mistrustful (sometimes also called distrustful) parties.  This interest was heightened by the
growing realisation that, by combining quantum protocols with 
relativistic signalling constraints, quite a wide
variety of tasks in mistrustful cryptography can be 
implemented with unconditional security.  Early examples
include relativistic bit commitment protocols
\cite{kentrel,kentrelfinite} that are provably secure
\cite{kentrelfinite} against all
classical attacks and against Mayers-Lo-Chau quantum attacks \cite{lomulti,lochauprl,mayersprl}.  
Other examples include the BHK quantum key distribution protocol based on no-signalling
(\cite{bhk} ; see also Ref.~\cite{BKP} for some further
  details and discussion) and later protocols significantly developing
  the idea ~\cite{AGM,AMP,ABGMPS,PABGMS,McKague,
    MRC,Masanes,HRW2,MPA,HR}, protocols for an interesting novel cryptographic
task, variable bias coin tossing \cite{vbct}, randomness expansion protocols (\cite{ColbeckThesis}, with a more complete presentation
in \cite{ckrandom}; \cite{PAMBMMOHLMM}) using untrusted devices, 
together with partial security results, and recent work on quantum
tagging (also called quantum position authentication) \cite{taggingpatent,malaney,chandranetal,kms,kcryptotagging,buhrmanetal,laulo}
and other forms of position-based
quantum cryptography. 

A very recent example, particularly relevant to the discussion of this
paper, is a simple new provably unconditionally secure protocol for
bit commitment \cite{bcsummoning} via securely transmitted qudits, which makes essential use of 
relativistic no-signalling constraints and of the properties of
quantum information, in particular the no-summoning theorem
(\cite{nosummoning}; see also \cite{otsummoning} for another
cryptographic application). 

Mistrustful classical cryptography
is relatively well understood.  
The relations between various important classical cryptographic primitives --- 
for example, coin tossing, bit commitment, the various equivalent versions of
oblivious transfer and secure multi-party computation --- have mostly
been established, along with some results on the 
composability of these primitives. 

There was initially some optimism that 
mistrustful quantum cryptography could be understood as a
straightforward generalisation of mistrustful 
classical cryptography.  On this view,
the role of the quantum cryptologist 
would be to investigate the possibility of secure quantum protocols   
which implement precisely the known classical primitives, with 
precisely the same composability properties.  However 
this ambition was arguably always misguided (see e.g. 
Ref. \cite{rudolph} for an early discussion) and was soon
frustrated. 
As we discuss below, the problem is that requiring a quantum 
protocol for a task to be unconditionally secure is generally
logically inconsistent with ideal classical cryptographic models
for that task.  In particular the superposition principle and the 
unitarity of quantum evolution are generally 
inconsistent with classically motivated definitions.  

\subsection{Classical certification} 
 
The introduction of relativistic protocols
adds another layer of complexity to questions about
what can and cannot be achieved by physics-based
cryptography.   As already mentioned, there are tasks for 
which one can prove that there are no unconditionally
secure non-relativistic protocols, but there are provably 
unconditionally secure relativistic protocols.  
As the history of non-relativistic quantum
cryptology already contains quite a few refuted conjectures
and subtle clarifications, it is perhaps no surprise that some 
rather basic questions about the possible scope of mistrustful
physics-based cryptography remain a source of some confusion 
to this day.   

This paper addresses one key point: the question of whether
{\it classical certification} is possible.  
That is: can a cryptographic protocol guarantee, based on physical principles alone,
to all the other parties that one party, Alice, is restricted so that (in
order to avoid being detected cheating) her quantum inputs must take
the form of pure states that are elements of a public fixed basis $\{ \ket{i}
\}_{i=1}^d$ of the appropriate $d$-dimensional input space ${ \cal H}_I$?
If there is such a guarantee, Alice is effectively
required to input classical information, since her input states can be
faithfully copied arbitrarily many times, either by her or any 
recipient, simply by measuring in the fixed basis and making copies of
the outcome states.   In other words, the protocol effectively {\it
  certifies} to the other parties that Alice is inputting only
classical information.  
If not, Alice is free to input not only superpositions $\sum_i a_i
\ket{i}$ of basis elements but also states in ${ \cal H}_I$
that are entangled with other systems ${\cal H}_A$ that she may use
elsewhere in the 
protocol, by creating states of the form 
$ \sum_i a_i \ket{i}_A \ket{i}_I $.   In other words, Alice is free to
input any quantum states lying in the appropriate input spaces.   

Classical classification would certainly be desirable in many
contexts: by taking these intrinsically quantum options away from
Alice one could ensure that the quantum protocol precisely replicates
a known classical task.  
However, we give here a simple argument to show that classical certification cannot be guaranteed
by quantum protocols for mistrustful cryptographic tasks. 
This argument applies both to non-relativistic protocols 
and to protocols using relativistic signalling constraints.
It is simpler than and supersedes an 
earlier argument applying to the
particular case of bit commitment \cite{akbccc}.

\section{A model of mistrustful relativistic quantum cryptography}

We model mistrustful quantum protocols in Minkowksi space as follows.  
The protocol involves a number $n \geq 2$  of
participating parties labelled $A, B , \ldots$.
Each party has a finite number of agents ($A_1 , \ldots , A_{n_A}$ and so on) in
secure laboratories; they trust everything inside their laboratories
and that all their operations within the laboratories are secure
against
eavedropping, but nothing outside the laboratories.
The representatives are linked by secure quantum channels, which
we can take to lie within the laboratories.\footnote{Alternatively,
we could take them to be teleportation channels, in which case they
can be jammed but not usefully eavesdropped.}  
 
We describe the protocol from the perspective of one party, 
say Alice, represented collectively by $A_1 , \ldots , A_{n_A}$. 
During the protocol the $A_i$ control space-time regions
$R_1 , \ldots , R_{n_A}$ that may be disconnected from one another 
but that are, we assume, each connected. 
The protocol defines {\it all} the actions that the $A_i$ collectively
should carry out, assuming Alice wishes to follow it honestly. 
This includes the generation and distribution of any inputs private
to Alice, and any entangled states required for any purpose
(communication, correlated inputs, $\ldots$).  This applies 
whether these inputs and entangled states need to be 
generated before data is exchanged with other parties, or during
the data exchange phase.  
The protocol fixes points $P_j$ within $R_i$ at which
{\it input states}, 
(which without loss of generality we take to be qubits $\ket{\psi_j}$), are generated 
from private data and prescribes how they are then propagated and processed.  
The protocol may also require $A_i$ at one or more points to choose at  
random an input state from a list $\ket{\psi_1} , \ldots , \ket{
  \psi_n }$; if so, 
it stipulates the relevant probabilities $p_1 , \ldots , p_n$. 

For any given spacelike section of $R_i$, the protocol stipulates a set of points in 
space-time on the boundary of $R_i$ at which $A_i$ must be
prepared to receive quantum states, another set from which she must be 
prepared to send quantum states, and a quantum network (which she may be
required to alter over time) within $R_i$ linking these sets.      

The protocol is supposed to be unconditionally secure -- i.e. to
have security based on the known laws of physics rather than any assumed
technological constraints.  In analysing the constraints on $A$ we
thus need to assume that each $A_i$ has
effectively unbounded quantum technology.  In particular, she 
can carry out quantum computations of arbitrary complexity effectively
instantaneously, send quantum states at light speed along error-free 
channels within the regions she controls, and store arbitrarily large
amounts of quantum information.  

The stage at which the protocol terminates may be pre-determined or may
be determined by collective computations carried out within the protocol.  In any case, we
assume it must terminate after a finite number of inputs, but we do not assume
there is any pre-determined bound on this number.\footnote{This is one
reason why we must allow the protocol to include instructions to each
party for generating and sharing inputs and entangled states among
their agents: any finite number of pre-generated inputs and shared entangled
states may be inadequate.}

The protocol may include security tests, which (without loss of
generality) we assume are defined by 
binary projective quantum measurements to be carried out by parties
after the rest of the protocol is complete.\footnote{A security test involving more
a general quantum measurement can always be written as one involving
a projective measurement on a larger quantum system.   A measurement
prescribed to be carried out during the protocol can always be 
postponed to the end of the protocol, again if necessary enlarging
the relevant quantum system.} 
These produce outputs, $1$ (``pass'') or $0$ (``fail''). 
We require that the protocol is {\it perfectly feasible}: if all parties are 
honest, then it will run to completion.   We also require that it is 
{\it perfectly reliable}: if all
parties honestly follow the protocol, then all security
tests always produce the outcome ``pass''.

Formally, then, the protocol prescribes for each party 
quantum algorithms to be run over prescribed quantum networks at each site, with prescribed
input and output channels and timings, together with
prescribed initial input data or random choices, which may be predistributed
via correlated states representing any required replication, either at the same site or at separated
sites. (For example, 
$a \ket{0}_{A_1} \ket{0}_{A_1} \ket{0}_{A_2} \ldots \ket{0}_{A_{N_A}} + 
b \ket{1}_{A_1} \ket{1}_{A_1} \ket{1}_{A_2} \ldots \ket{1}_{A_{N_A}}$
represents a random input bit, with $p(0) = |a|^2$, replicated so that
two input copies are available at $A_1$ and one at all other sites.)

\section{Classical certification is impossible}

Suppose now that we have a protocol which guarantees classical
certification for Alice's inputs.  Consider a single classically certified bit input 
into a protocol by $A_i$.  Without loss of generality
we suppose the protocol allows either classical bit value as input
(otherwise the input is trivial).  
If $A_i$ chooses to input the state $\ket{0}$, representing the
classical bit $0$, she prepares $\ket{0}$ and 
inserts it at the appropriate point into her quantum network.
Similarly, to input $\ket{1}$, representing the
classical bit $1$, she prepares and inserts $\ket{1}$. 

Now suppose that she chooses instead to prepare the state $a \ket{0}
 + b \ket{1}$.  By
assumption, the probability of any security measurement $P$ producing
outcome ``fail'' is zero in the first two cases.  Hence, by linearity,
the probability of ``fail'' is zero in the third case.  This
contradicts the assumption that the protocol guaranteed classical
certification of the bit.

Similarly, if the protocol requires $A_i$ to a input a state chosen from
the ensemble $ \{ \ket{ \psi_i } ; p_i \}$, she can instead prepare
a state of the form $\sum_i p_i^{1/2} \ket{i}_{AS} \ket{ \psi_i }_I$,
where the $\ket{i}_{AS}$ form an orthonormal basis of an ancillary
system that she stores, and then input the $I$ system.   
Since no measurement can distinguish between proper and improper 
mixtures represented by the same density matrix, and the probability
of any security test producing ``fail'' is zero if Alice follows the
protocol faithfully, it must also be zero if she deviates in this way.  

Obviously, these arguments extend immediately to inputs of states
of arbitrary finite dimension.  
Classical certification of Alice's inputs (whether defined by private
data or randomly chosen) is thus impossible, as claimed.

\section{Why classical certification cannot generally be enforced by measurement}

One might possibly be tempted to think that (without contradicting
the above proof) a property
operationally equivalent to classical certification 
can easily be guaranteed, since
even if one party inputs a superposition of bits into a protocol,
any other party can collapse the superposition by carrying out 
a measurement on the input in the computational basis.  

This is generally incorrect.  In general, the parties input bits into 
their own quantum computers, which process the quantum data,
along with data received earlier in the protocol, before 
sending appropriate subsets to another party or parties. 
Consider a single input qubit, and two possible orthogonal input
states, $\ket{0}$ and $\ket{1}$.  Although the corresponding 
output states must be orthogonal, the reduced density matrices
for the corresponding states sent on to the other parties, $\rho_0$ and $\rho_1$, need
not necessarily be.   Also, whether or not they are orthogonal, the receiving party
may not necessarily know the measurement basis
which (perfectly or optimally) distinguishes them.

\section{Discussion}

We have given a simple general argument against the possibility
of physically guaranteed certificates of classicality for mistrustful
cryptographic protocols.  This addresses a point which seems to have caused some 
confusion.  If we were to require that mistrustful 
quantum protocols should follow ideal  
classical definitions precisely, as has sometimes been suggested
in the literature, then in particular we would have to 
require mistrustful quantum protocols to guarantee
classical certification of their inputs, and this would either
trivialise or exclude many of the most interesting questions in
mistrustful quantum cryptology.  

For example, if we were to require --
as a matter of definition -- that any quantum bit commitment protocol
must guarantee classical certification of the committed bit, we would
not need Mayers' and Lo-Chau's celebrated and elegant demonstrations
\cite{mayersprl,lochauprl} of the impossibility of non-relativistic
quantum bit commitment: the one-line proof given in this paper would
suffice.   

We hope this small clarification
will help focus attention on attainable quantum cryptographic security
criteria. 

\vskip5pt \leftline{\bf Acknowledgments}

I thank Roger Colbeck, Hoi-Kwong Lo, J\"orn M\"uller-Quade and Dominique Unruh 
for helpful comments, and acknowledge partial support from the 
Cambridge-MIT Institute, the project PROSECCO (IST-2001-39227) of the IST-FET 
programme of the EC, and the Perimeter Institute.   Research at
Perimeter Institute is supported by the Government of Canada through Industry Canada and
by the Province of Ontario through the Ministry of Research and Innovation.
%
%


\begin{thebibliography}{99} 
\bibitem{wiesner}
S.~Wiesner, Conjugate coding, {\it SIGACT News} {\bf 15(1)} 78-88 (1983).  
\bibitem{BBeightyfour}
C. H. Bennett and G. Brassard, Quantum cryptography:
Public-key distribution and coin tossing, {\it Proceedings of the
International Conference on Computers, Systems and Signal Processing} (IEEE,
 New York, 1984), pp.~175-179.
\bibitem{ekert}
A.~Ekert, Quantum cryptography based on Bell's theorem, 
{\it Phys. Rev. Lett.} {\bf 67} 661-663 (1991).   

\bibitem{kentrel}
A.~Kent, Unconditionally secure bit commitment,
Phys. Rev. Lett. {\bf 83} 1447-1450 (1999).
\bibitem{kentrelfinite}
A.~Kent, Secure classical bit commitment using fixed capacity communication
channels, 
Journal of Cryptology, {\bf 18}, 313-335 (2005). 
\bibitem{bhk}
J.~Barrett, L.~Hardy and A.~Kent, 
No Signalling and Quantum Key Distribution
Phys. Rev. Lett. {\bf 95}, 010503 (2005).  

\bibitem{BKP}
\bibinfo{author}{Barrett, J.}, \bibinfo{author}{Kent, A.} \&
  \bibinfo{author}{Pironio, S.}
\newblock \bibinfo{title}{Maximally non-local and monogamous quantum
  correlations}.
\newblock \emph{\bibinfo{journal}{Physical Review Letters}}
  \textbf{\bibinfo{volume}{97}}, \bibinfo{pages}{170409}
  (\bibinfo{year}{2006}).

\bibitem{AGM}
\bibinfo{author}{Ac\'in, A.}, \bibinfo{author}{Gisin, N.} \&
  \bibinfo{author}{Masanes, L.}
\newblock \bibinfo{title}{From {B}ell’s theorem to secure quantum key
  distribution.}
\newblock \emph{\bibinfo{journal}{Physical Review Letters}}
  \textbf{\bibinfo{volume}{97}}, \bibinfo{pages}{120405}
  (\bibinfo{year}{2006}).

\bibitem{AMP}
\bibinfo{author}{Ac\'in, A.}, \bibinfo{author}{Massar, S.} \&
  \bibinfo{author}{Pironio, S.}
\newblock \bibinfo{title}{Efficient quantum key distribution secure against
  no-signalling eavesdroppers}.
\newblock \emph{\bibinfo{journal}{New Journal of Physics}}
  \textbf{\bibinfo{volume}{8}}, \bibinfo{pages}{126} (\bibinfo{year}{2006}).

\bibitem{ABGMPS}
\bibinfo{author}{Acin, A.} \emph{et~al.}
\newblock \bibinfo{title}{Device-independent security of quantum cryptography
  against collective attacks}.
\newblock \emph{\bibinfo{journal}{Physical Review Letters}}
  \textbf{\bibinfo{volume}{98}}, \bibinfo{pages}{230501}
  (\bibinfo{year}{2007}).

\bibitem{PABGMS}
\bibinfo{author}{Pironio, S.} \emph{et~al.}
\newblock \bibinfo{title}{Device-independent quantum key distribution secure
  against collective attacks}.
\newblock \emph{\bibinfo{journal}{New Journal of Physics}}
  \textbf{\bibinfo{volume}{11}}, \bibinfo{pages}{045021}
  (\bibinfo{year}{2009}).

\bibitem{McKague}
\bibinfo{author}{McKague, M.}
\newblock \bibinfo{title}{Device independent quantum key distribution secure
  against coherent attacks with memoryless measurement devices}.
\newblock \bibinfo{howpublished}{e-print \url{arXiv:0908.0503}}
  (\bibinfo{year}{2009}).

\bibitem{MRC}
\bibinfo{author}{Masanes, L.}, \bibinfo{author}{Renner, R.},
  \bibinfo{author}{Christandl, M.}, \bibinfo{author}{Winter, A.} \&
  \bibinfo{author}{Barrett, J.}
\newblock \bibinfo{title}{Unconditional security of key distribution from
  causality constraints.}
\newblock \bibinfo{howpublished}{e-print \url{quant-ph/0606049v4}}
  (\bibinfo{year}{2009}).

\bibitem{Masanes}
\bibinfo{author}{Masanes, L.}
\newblock \bibinfo{title}{Universally composable privacy amplification from
  causality constraints}.
\newblock \emph{\bibinfo{journal}{Physical Review Letters}}
  \textbf{\bibinfo{volume}{102}}, \bibinfo{pages}{140501}
  (\bibinfo{year}{2009}).

\bibitem{HRW2}
\bibinfo{author}{H\"anggi, E.}, \bibinfo{author}{Renner, R.} \&
  \bibinfo{author}{Wolf, S.}
\newblock \bibinfo{title}{Quantum cryptography based solely on {B}ell's theorem}.
\newblock In \bibinfo{editor}{Gilbert, H.} (ed.)
  \emph{\bibinfo{booktitle}{Proceedings of the 29th Annual International
  Conference on the Theory and Applications of Cryptographic Techniques
  (Eurocrypt'10)}}, \bibinfo{pages}{216--234} (\bibinfo{publisher}{Springer},
  \bibinfo{year}{2010}).
\newblock \bibinfo{note}{Also available as \url{arXiv:0911.4171}}.

\bibitem{MPA}
\bibinfo{author}{Masanes, L.}, \bibinfo{author}{Pironio, S.} \&
  \bibinfo{author}{Ac\'in, A.}
\newblock \bibinfo{title}{Secure device-independent quantum key distribution
  with causally independent measurement devices}.
\newblock \bibinfo{howpublished}{e-print \url{arXiv:1009.1567}}
  (\bibinfo{year}{2010}).

\bibitem{HR}
\bibinfo{author}{H\"anggi, E.} \& \bibinfo{author}{Renner, R.}
\newblock \bibinfo{title}{Device-independent quantum key distribution with
  commuting measurements}.
\newblock \bibinfo{howpublished}{e-print \url{arXiv:1009.1833}}
  (\bibinfo{year}{2010}).

\bibitem{vbct}
Variable Bias Coin Tossing, R.~Colbeck and A.~Kent 
Phys. Rev. A {\bf 73}, 032320 (2006).
\bibitem{ColbeckThesis}
\bibinfo{author}{Colbeck, R.}
\newblock \emph{\bibinfo{title}{Quantum and Relativistic Protocols For Secure
  Multi-Party Computation}}.
\newblock Ph.D. thesis, \bibinfo{school}{University of Cambridge}
  (\bibinfo{year}{2007}).
\newblock \bibinfo{note}{Also available as \url{arXiv:0911.3814}.}

\bibitem{PAMBMMOHLMM}
\bibinfo{author}{Pironio, S.} \emph{et~al.}
\newblock \bibinfo{title}{Random numbers certified by {B}ell's theorem}.
\newblock \emph{\bibinfo{journal}{Nature}} \textbf{\bibinfo{volume}{464}},
  \bibinfo{pages}{1021--1024} (\bibinfo{year}{2010}).

\bibitem{ckrandom}
R.~Colbeck and A.~Kent, 
Private Randomness Expansion With Untrusted Devices,
J. Phys. A {\bf 44} 095305
(2011). 

\bibitem{taggingpatent}
A.~Kent, R.~Beausoleil, W.~Munro and T.~Spiller,
{\it Tagging Systems}, 
US patent US20067075438 (2006).
\bibitem{malaney}
R.~Malaney, Phys. Rev. A {\bf 81}, 042319 (2010) ( arXiv:1003.0949 ); 
R.~Malaney, arXiv:1004.4689 (2010). 
\bibitem{chandranetal}
N.~Chandran et al., arXiv:1005.1750 (2010).   
\bibitem{kms}
A.~Kent, W.~Munro and T.~Spiller, 
Quantum Tagging: Authenticating Location via Quantum Information and Relativistic Signalling Constraints, 
arxiv:1008.2147 (2010), 
to appear in Phys. Rev. A.   
\bibitem{kcryptotagging}
A.~Kent,
Quantum Tagging with Cryptographically Secure Tags,
arxiv:1008.5380. 
\bibitem{buhrmanetal}
H.~Buhrman et al., arXiv:1009.2490v3 (2010). 
\bibitem{laulo}
H.K.~Lau and H.K.~Lo, 
Insecurity of position-based quantum cryptography protocols against entanglement attacks, 
Phys. Rev. A 83, 012322 (2011).

\bibitem{bcsummoning}
A.~Kent,
Unconditionally Secure Bit Commitment with Flying Qudits,
arXiv:1101.4620 (2011). 
\bibitem{nosummoning}
A.~Kent, 
A No-summoning theorem in Relativistic Quantum Theory,
arXiv:1101.4612 (2011). 
\bibitem{otsummoning}
A.~Kent,
Location-Oblivious Data Transfer with Flying Entangled Qudits, 
arxiv:1102:2816 (2011).
\bibitem{rudolph} 
T.~Rudolph, The Laws of Physics and Cryptographic Security, quant-ph/0202143. 
\bibitem{akbccc}
A.~Kent, Unconditionally Secure Commitment of a Certified Classical Bit is Impossible,
Phys. Rev. A {\bf 61}, 042301 (2000).
 \bibitem{lomulti}
H.-K.~Lo,  Insecurity of Quantum Secure Computations, 
Phys. Rev. A {\bf 56} (1997) 1154.
\bibitem{lochauprl}
H.-K.~Lo and H.~Chau, Is quantum bit commitment really
possible?,  Phys. Rev. Lett. {\bf 78}
3410-3413  (1997).
\bibitem{mayersprl}
D.~Mayers, Unconditionally secure quantum bit commitment
is impossible, Phys. Rev. Lett. {\bf 78} 3414-3417 (1997). 
\end{thebibliography}
\end{document}